\documentclass[useAMS,usegraphicx,usenatbib]{mn2e}

\newcommand{\GJ}[1]{\ensuremath{#1_\textrm{\tiny GJ}}}

\newcommand{\PC}[1]{\ensuremath{#1_\mathrm{pc}}}    

\newcommand{\NS}[1]{\ensuremath{#1_{\textrm{\tiny NS}}}}    

\newcommand{\Wmd}{\protect{\ensuremath{W_\mathrm{md}}}} 
\newcommand{\RLC}{\ensuremath{R_\mathrm{LC}}}     

\newcommand{\A}[1]{\protect{\ensuremath{#1_\mathrm{A}}}} 
\newcommand{\B}[1]{\protect{\ensuremath{#1_\mathrm{B}}}} 

\graphicspath{{./figures/}}

\title[Nulling and mode changing in pulsars]{A model for nulling and mode changing in pulsars}

\author[A.~N.~Timokhin]{A.~N.~Timokhin$^{1,2}$\thanks{E-mail: atimokhin@berkeley.edu}\\
  $^{1}$Astronomy Department, University of California at Berkeley,
  601 Campbell Hall, Berkeley, CA 94720, USA\\
  $^{2}$Sternberg Astronomical Institute, Universitetskij pr. 13,
  Moscow 119992, Russia}

\begin{document}

\date{Received ; in original form }

\pagerange{\pageref{firstpage}--\pageref{lastpage}} \pubyear{2010}

\maketitle

\label{firstpage}

\begin{abstract}
  We propose that in some pulsars the magnetosphere has different
  states with different geometries or/and different distributions of
  currents, it occasionally switches between them.  These states have
  different spindown rates and emission beams, in some of the states
  no radioemission is produced at all.  Switching into a different
  state manifests as a mode change when we see different parts of the
  emission beam or the beams in different states have significantly
  different geometries, it manifests as nulling when we either miss
  the new beam or no radioemission is generated in the new state.  We
  show that modest variations in the beam shape can be accompanied by
  large variations in the pulsar spindown rate $W$ -- the dependence
  of $W$ on the opening angle of the emission beam $\alpha$ can be as
  strong as $W\propto{}\alpha^4$.  We speculate about physical
  mechanisms which may cause reconfiguration of the magnetosphere.
\end{abstract}


\section{Introduction}

Rotationally powered pulsars are observed mostly in radio band because
of the very high sensitivity of radio telescopes.  Pulsar
radioemission shows very rich set of phenomena.  Examples of such
phenomena are mode changing and nulling, which are observed in many
pulsars.  There are hints that mode changing and nulling may be
manifestations of the same phenomenon \citep[e.g.][]{Wang2007}.  It is
not clear yet whether nulling and/or mode changes are 'uniform'
phenomena, i.e. whether there is a single class of reasons why pulsars
null or/and change modes.  There are nulling pulsars which are quiet
for only a few periods and there are pulsars which spend very long
time in each of the state.  There are pulsar for which the emission
intensity during nulls drops at least by several orders of magnitude
or switches off completely so that they cannot be detected even in a
deep search, while for some pulsars nulls seems to be just mode(s)
with low emission intensity \citep[e.g.][]{Wang2007,Biggs1992}.

It was not clear for a long time whether nulling is due to
'micro-physics' of the radioemission mechanism or changes in the
pulsar magnetosphere as a whole play a role.  The discovery of two
nulling pulsars PSR B1931+24 and PSR J1832+0029 with different
spindown rates in ON and OFF phases \citep{Kramer2006,Lyne2009}
provides strong evidence that at least some subclass of nulling may be
related to the global properties of pulsars and not merely a property
of radioemission mechanism(s) alone.  In this short paper we will
focus on pulsars where global processes involving changes in the whole
magnetosphere seems to be at play.  We propose a qualitative model
which could explain nulling and mode changing as manifestation of the
same phenomenon and account for the observed variation of the spindown
rate.

\section{The model}

The radioemission of a pulsar is negligible in the overall energy
budget, usually constituting a very tiny fraction of the pulsar
spindown rate, less then $10^{-3}$.  Most of the energy flux is
carried away by the relativistic pulsar wind and does not reveal
itself in the emission.  Therefore, merely the fact whether the
radioemission is produced or for some reason is switched off or
strongly suppressed cannot change the pulsar spindown such that it
would be measurable.  However, there are at least two pulsars -- PSR
B1931+24 and PSR J1832+0029 -- which exhibit nulling and at the same
time their spindown rates are substantially less when they are
non-visible \citep{Kramer2006,Lyne2009}.  In the case of PSR B1931+24
the difference in the spindown rate between the ON and OFF modes is
about 50\%.  Nulling in these pulsars takes an extreme form, they
spend a very long time in both state (days), whereas many other
nulling pulsars switch off for much shorter times -- hours or less
\citep[e.g.][]{Biggs1992,Wang2007}.  Taking into account the smallness
of the pulsar period derivatives, $\dot{P}\sim{}10^{-15}$, it is very
hard to detect variations of spindown rate between modes lasting only
hundreds or thousands of pulsar periods, even if these variations are
large.  We may speculate that other nulling pulsars, or at least some
of them, can have non-negligible variations of the spindown rate too,
but they are difficult to measure and are not detected.

The behavior of PSR B1931+24 and PSR J1832+0029 hints that nulling, at
least in some pulsars, is a manifestation of some global process which
changes the spindown rate of pulsar.  This global process should
involve changes in the structure of the pulsar magnetosphere.  The
most popular radiopulsar model, introduced by \citet{GJ}, implies that
the pulsar magnetosphere is filled with plasma, and accelerating
electric field is present only in geometrically very small regions --
in the pulsar polar cap and in regions around and inside the current
sheet carrying the bulk of the return current.  The rest of the
magnetosphere should be force-free.  In the force-free magnetosphere
the energy losses of pulsar are \emph{determined} by the configuration
of the magnetosphere.  Within this configuration we understand both
the geometrical relation between the open and closed magnetic field
line zones as well as the distribution of the current density in the
open field line zone.  In other words, in the force-free model changes
in the pulsar spindown rate are possible only if the configuration of
the magnetosphere changes.  Recently the force-free pulsar
magnetosphere model has been studied in great detail
\citep[e.g.][]{CKF,Gruzinov:PSR,Contopoulos05,Timokhin2006:MNRAS1,Timokhin2007:MNRAS2,
  Spitkovsky:incl:06}.  For an aligned rotator it was shown that there
exists: i) stationary magnetospheric configurations with different
sizes of the closed magnetic field line zone
\citep{Timokhin2006:MNRAS1}; and ii) configurations with different
current density distributions in the open field line zone
\citep{Contopoulos05,Timokhin2007:MNRAS2,timokhin::PSREQ_london/2006}. These
configurations have different spindown rates.

Having the above described facts, hints and speculations in mind we
make the following conjecture.  Magnetospheres of some pulsars at some
stages of their evolution can have sets of quasi-stable states with
different sizes of the closed field line zone or different current
density distributions in the open field line zone, or both.
Occasionally a magnetosphere switches between these states.

Currently there is no reliable theory which can explain pulsar
radioemission.  The radioemission is coherent and it seems that
conditions for generation of coherent emission in pulsar magnetosphere
may be not always fulfilled -- at least there are pulsars with periods
and magnetic fields typical for radioemitting pulsars which are
visible in gamma-rays but show no detectable radioemission
\citep{FERMI_PSR_Catalog2010}.  If, at least for some range of pulsar
parameters, the conditions for generation of coherent emission are
very sensitive to the properties of the magnetospheric plasma -- like
current and plasma densities or/and their gradients -- then it can be
that in some of the magnetospheric states the coherent emission is not
generated at all or is generated at different places.  Each state then
has unique radioemission properties and switching between different
magnetospheric states will manifest as mode changing or nulls.

Another reason for changes of the pulsar mean profile can be due to
self-similar changes of the emission beam when the radioemission zone
shrinks or expands, i.e. radioemission is generated in the same way in
all states but the beam scales self-similarly with the size of the
emission zone.  Then the emission beam in some states is wider than in
another.  Depending on the orientation of our line of sight relative
to the emission beam we can see either mode change or null depending
on whether our line of sight still crosses the narrower emission beam
or not.

We note that in the proposed model the magnetosphere is always filled
with plasma and qualitatively works in the same way, i.e. particles
are accelerated, gamma-rays are emitted etc.  But changes in the beam
pattern due to current redistribution or/and shrinking of the
corotating zone are always accompanied by changes in the spindown
rate.  Below we try to quantify the proposed model.  We defer
speculations about reasons causing such behavior of the magnetosphere
to Discussion.

We do not know what the conditions for coherent emission of radio
waves in the pulsar magnetosphere are.  Hence, we can not make any
quantitative statements about what changes in the magnetospheric
configuration can cause failure of the emission mechanism or what is
the necessary condition for large changes in the beam pattern when the
magnetospheric configuration changes.  However, we can estimate how
the emission beam changes when the emission patterns in all
configurations are similar -- when nulling/mode changes are due to
pure geometrical effects.  Even if contraction/expansion of the
emission beam takes place only in some nulling/mode changing pulsars,
detection of a correlation between the width of the mean profile and
the spindown rate in a mode-changing pulsar could be a strong argument
in favor of the proposed model.  On the other hand, for pulsars where
nulling/mode changes cannot be (fully) explained by
contraction/expansion of the emission beam, such estimates could
provide useful limits on variation of the spindown rate.

For simplicity we consider only the case of an aligned rotator with
a dipole magnetic field.  The main point we wish to demonstrate with
these crude estimations is that \emph{large} variations of the spindown
rate can be accompanied by \emph{modest} changes in the geometry of
the radio beam, or, reversing the statement, if changes of the beam
width due to reconfiguration of the force-free magnetosphere are
observed, then changes in the spindown rate would be large and could
be detected.

\subsection{Shrinking-expanding corotating zone}
\label{sec:shrink-expand-corot}

\begin{figure}
  \begin{center}
    \includegraphics[clip,width=0.75\columnwidth]{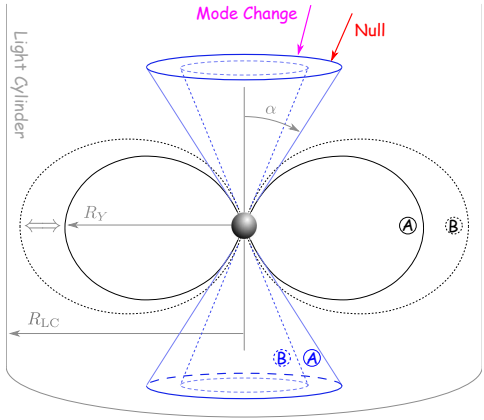}
  \end{center}
  \caption{Schematic view of pulsar magnetosphere when the closed
    field line zone changes its size.  Radioemission beams are shown
    by the blue lines, the last closed field lines -- by the black
    lines.  The lines of sight for which pulsar will demonstrate mode
    change and null are shown by magenta and red arrows respectively.}
  \label{fig:magn_change_x0}
\end{figure}

Let us assume that a pulsar magnetosphere has different states with
different sizes of the corotating zone, the zone with closed magnetic
field lines.  Let us consider two states: state A with the corotating
zone smaller than that in the state B (see
Fig.~\ref{fig:magn_change_x0}).  The size of the polar cap, limited by
the last closed field line is smaller in state B.  The current density
distribution in the polar cap of pulsar for such states do not differ
dramatically (see Fig.~(5) in \citet{Timokhin2006:MNRAS1}).  It is
naturally to assume that all processes in regions not far from the
neutron star (NS) for small changes in the size of the polar cap
behave similarly, i.e. the beaming pattern in different configurations
can be approximately described by a shrinking/contraction proportional
to the changes in the size of the polar cap.

If radio emission is generated close to the NS, then the emission beam
in state A has a larger opening angle.  When the pulsar switches from
state A into state B, depending on our line of sight, we see either
different parts of the emission cone, what might result in a mode
change, or we miss the emission cone entirely and conclude that pulsar
is in the 'null' state.  Configurations with a smaller corotating zone
have more open magnetic field lines and, hence, have higher energy
flux to infinity.  Therefore, during the ON state the spindown rate of
pulsar is higher.

If emission is directed along magnetic field lines, then the opening
angle of the emission cone $\alpha$ is
\begin{equation}
  \label{eq:alpha}
  \alpha = \theta + \arctan\left(B_\theta/B_r\right) \simeq
  3\,\theta/2 \propto \PC{\theta}
\end{equation}
where $\theta$ is the colatitude of the emission zone, $B_\theta,B_r$
are components of the magnetic field in spherical coordinates, and
$\PC{\theta}$ is the colatitude of the polar cap boundary.  The last
step in eq.~(\ref{eq:alpha}) comes from our assumption about
similarities of emission zones in different configurations.  Energy
losses of the aligned rotator depend on the size of the corotating
zone as (eq.~(62) in \citet{Timokhin2006:MNRAS1})
\begin{equation}
  \label{eq:W}
  W \simeq \Wmd \left(\frac{R_Y}{\RLC}\right)^{-2} 
\end{equation}
where $R_Y$ is the actual size of the corotating zone, $\RLC$ is the
size of the light cylinder. $\Wmd$ are the magnetodipolar energy losses
$\Wmd = B_0^2 \NS{R}^6\NS{\Omega}^4/4c^3$,
where $B_0$ is the magnetic field in the polar cap, $\NS{R}$ is the NS
radius, $\NS{\Omega}$ is the pulsar angular velocity, $c$ is the speed
of light.  For dipolar magnetic field
$\PC{\theta}\simeq\sqrt{\NS{R}/R_Y}$.  Expressing $R_Y$ through
$\PC{\theta}$ we get
\begin{equation}
  \label{eq:w_theta_pc}
  W \simeq \Wmd \left(\frac{\RLC}{\NS{R}}\right)^2 \PC{\theta}^4\,.
\end{equation}
Taking into account relation~(\ref{eq:alpha}) we finally get
\begin{equation}
  \label{eq:w_alpha__r_y}
  W \propto \alpha^4\,.
\end{equation}
We see that reconfiguration of the magnetosphere can indeed cause much
smaller changes in the beam opening angle than in the spindown rate,
and so small changes in the emission geometry can be accompanied by
measurable changes of the spindown rate.

\subsection{Changing current density distribution}
\label{sec:chang-curr-dens}

Let us now consider the case, when the polar cap size is nearly the
same ($R_Y$ does not change), but instead there are magnetospheric
states with different current density distributions in the open field
line zone.  A model of such type was qualitatively discussed by
\cite{Arons1983a} in relation to mode changing in PSR B0809+74.

The force-free magnetosphere starts at the top of the cascade zone in
the polar cap, above the pair formation front, where there is enough
plasma to short out the accelerating electric field.  The angular
velocity in the magnetosphere is constant along magnetic field lines.
Each open magnetic field line rotates with the angular velocity
$\Omega_F$ it has at the top of the cascade zone in the polar cap.
$\Omega_F$ depends on the distribution of the potential drop in the
cascade zone $V_{||}$ -- the potential drop between the base of the
force-free magnetosphere and the NS surface.  In the force-free
magnetosphere the current density distribution is set by the
requirement that the open magnetic field lines pass smoothly through
the light cylinder.  For magnetospheric states with different
distributions of the accelerating potential $V_{||}$ the distributions
of $\Omega_F$ will be also different, and so will be the positions and
the shapes of the light cylinder.  Hence, such states will have
different current density distributions in the open field line zone as
well.

Pulsar energy losses are given by (cf. eq.~(40) in
\citet{Timokhin2007:MNRAS2})
\begin{equation}
  \label{eq:W_I_Omega}
  W \sim \frac{J \Omega_F}{2\pi{}c}\PC{\Phi}\,.
\end{equation}
where $J$ is the total current flowing through the polar cap,
$\PC{\Phi}=B_0\pi\PC{r}^2$ is the total magnetic flux in the open
field line zone, $\PC{r}$ is the polar cap radius.  The dependence of
$J$ on $\Omega_F$ could be approximated by that dependence in the
magnetosphere of a rotating split-monopole $J\propto\Omega_F$
\citep{Blandford/Znajek77}.  And so we have for the energy losses
\begin{equation}
  \label{eq:W_I}
  W \propto J^2\,,
\end{equation}
the spindown rate is proportional to the square of the total electric
current flowing through the polar cap, a dependence typical for power
dissipated in an electric circuit.

To change the spindown rate the total current must change; small local
changes in the current density which do not change the total current
cannot account for changes in $W$.  As a toy model let us consider two
states A and B, where the shape of the current distribution remains
the same but the current density $j$ in the state B is less than that
in the state A by some factor $\eta$ slightly less than $1$,
\begin{equation}
  \label{eq:jA_jB}
  \B{j}=\eta\,\A{j}\,.
\end{equation}
The total current changes then as $\B{J}=\eta\A{J}$, and for changes
in the energy losses we have
\begin{equation}
  \label{eq:W_eta_I}
  W\propto{}\eta^2\,.
\end{equation}

There is no reliable model for pulsar radioemission.  We can only
speculate what sets the pattern for the emission beam.  Let us assume
that in state B the emission zone shifts to the field line where the
current density is equal to the current density which flowed through
the emission zone in state A, i.e.
\begin{equation}
  \label{eq:jA_em__jB_em}
  \B{j}(\B{\theta})=\A{j}(\A{\theta})\,.
\end{equation}
This toy model results in changes of the beam geometry similar to that
shown in Fig.~\ref{fig:magn_change_x0}.

The current density in the polar cap of the aligned rotator with
negligible differential rotation of the open magnetic field lines is
close to the \citet{Michel73} current density
\citep{Timokhin2006:MNRAS1}
\begin{equation}
  \label{eq:j_michel}
  j = \GJ{j} \left(1- \theta^2/\PC{\theta}^2 \right)\,,
\end{equation}
where $\GJ{j}$ is the Goldreich-Julian current density.  Using
relations~(\ref{eq:jA_jB}) and (\ref{eq:jA_em__jB_em}) with the
current density given by eq.~(\ref{eq:j_michel}), we get the following
expression for variation of the emission zone colatitude in the case
of small changes in the current density
\begin{equation}
  \label{eq:delta_theta/theta}
  \frac{\delta\theta}{\theta}\simeq
  \frac{1}{2}(\eta-1)
  \frac{1-\theta^2/\PC{\theta}^2}{\theta^2/\PC{\theta}^2}\,.
\end{equation}
From eq.~(\ref{eq:W_eta_I}) we have $\delta{}W/W = 2(\eta-1)$, and
small variation of the energy losses are related to variation of the
emission zone colatitude as
\begin{equation}
  \label{eq:delta_W__delta_theta}
   \frac{\delta{}W}{W} \simeq 4\zeta \frac{\delta\theta}{\theta}\,,
\end{equation}
where $\zeta=(\theta^2/\PC{\theta}^2)/(1-\theta^2/\PC{\theta}^2)$.  If
the emission zone is close to the edge of the polar cap,
$\theta>\PC{\theta}/\sqrt{2}\simeq{}0.71\,\PC{\theta}$, then
$\zeta>1$.  Taking into account the relation between the colatitude of
the emission zone and the beam opening angle~(\ref{eq:alpha}) we get
the power law
\begin{equation}
  \label{eq:W_alpha_j}
   W\propto{}\alpha^{4\zeta}\,,
\end{equation}
which is valid for small variations of $W$ and $\alpha$.  We see that
in this case changes in the spindown rate could be even larger than in
the case of a changing corotating zone discussed before.  Although the
model considered here is very simple, it demonstrates nevertheless
that the dependence of the energy losses on the beaming angle can be
rather strong.

\section{Discussion}

Let us summarize the reasoning behind the proposed model.  It seems
that the force-free model agrees well with observations: pulses are
narrow, pointing to smallness of the regions with accelerating
electric field; as it is inferred from observations of pulsar wind
nebulae, there should be more than enough plasma in the magnetosphere
to screen the accelerating electric field.  If pulsar magnetosphere is
essentially force-free, then changes of the spindown rate are possible
only if the configuration of the magnetosphere changes.  So, in the
case of PSR~B1931+24 and PSR~J1832+0029 we must conclude that we have
to do with changes in the magnetosphere configuration, or the
force-free model fails.  Different configurations should have
different emission-beam geometries.

We do not suggest that all nulling pulsars switch between states with
different emission beams, neither that changes in the emission beam
could be always attributed to the pure geometrical effect of
shrinking/expansion; the latter can work only for pulsars where the
line of sight crosses the outer parts of the emission beam. 
In the case of PSR~B1931+24 the line of sight seems to make central
traverse of the beam (Rankin, private communication), and so the
simple geometrical models of
Sec.~\ref{sec:shrink-expand-corot},~\ref{sec:chang-curr-dens} do not
work for it. In the frame of the proposed model nulls in this pulsar
should be then attributed to cessation of the radioemission due to
changed parameters of magnetospheric plasma in the OFF state.
But if pulsar magnetosphere can switch between different states,
shrinking/expansion of the emission beam can take place in some
pulsars -- at least in the mode changing pulsar PSR~B0943+10 with
tangential sightline trajectory the modes have different widths of the
mean profile \citep{RankinSuleymanova2006}.  The point we wish to
emphasize here is that reconfiguration of the magnetosphere can result
in modest changes in the emission beam and in substantial changes in
the spindown rate at the same time, i.e. the emission beams in
different states can look similar while the energy loss rate changes a
lot.

An important consequence of the proposed model is that there must be
mode changing pulsars showing substantial variations of spindown rate,
so to say, the mode-changing ``twins'' of PSR B1931+24 and PSR
J1832+0029.  If the magnetospheres can have different states,
switching between the states should manifest as mode changing or
nulls.  The spindown rate variation in the above mentioned nulling
pulsars strongly suggests that such states exist.  Hence, there also
must be pulsars where states with different spindown rates manifest as
different emission modes.  The proposed model also implies that
nulling and mode changing pulsars should be among the pulsars with
high timing noise.

Let us now discuss what could be the underlying physical mechanisms
for the proposed model.  It is not a question whether for a given
pulsar period and magnetic field strength different configurations of
the magnetosphere are admitted or not.  At least for the aligned
rotator it has been shown explicitly that such configurations exist,
and we see no reasons why this can not be the case for the inclined
rotator.  The real question is why the magnetosphere can have a set of
\emph{meta-stable} configurations, i.e. configurations where the
magnetosphere can stay for times much longer than the pulsar period.

Although the properties of the force-free magnetosphere are quite well
known, we cannot say what is the configuration of a pulsar
magnetosphere. The reason for this is that the configuration of the
magnetosphere depends on the ``boundary conditions'' -- physics of the
polar cap cascade zone and physics of magnetic reconnection in the
current sheet, especially at the so-called ``Y''-point where the
current sheet of the outer magnetosphere merges with that along the
boundary between open and closed magnetic field lines.  The
distribution of the potential drop in the polar cap cascade zone sets
the angular velocity of the open magnetic field lines $\Omega_F$
\citep[e.g.][]{Contopoulos05,Timokhin2007:MNRAS2,timokhin::PSREQ_london/2006},
the reconnection rate at ``Y''-point might influence the size of the
of the corotating zone \citep[e.g.][]{Contopoulos2006,Bucciantini06}.
The physics of both of these regions is poorly understood, so we can
only speculate about what properties the whole system consisting of
the force-free magnetosphere + the cascade zone + the current sheet
can have.  Here we suggest one possible scenario for existence of
several meta-stable magnetosphere configurations.

The total energy of the magnetosphere of the aligned rotator with
$\Omega_F\equiv\NS{\Omega}$ monotonically depends on the size of the
corotating zone, it decreases with the increase of $R_Y$
\citep{Timokhin2006:MNRAS1}.  The total energy of the split-monopole
magnetosphere with different current densities monotonically increases
with the increase of the total current \citep{Timokhin2007:MNRAS2}.
It can be that the combining effect of different current density
distributions and different sizes of the corotating zone, especially
in the inclined rotator%
\footnote{In the time-dependent numerical simulations of the inclined
  rotator magnetosphere by \citet{Spitkovsky:incl:06} and
  \citet{Kalapotharakos2009} constant $\Omega_F\equiv\NS{\Omega}$ and
  fast reconnection rate at Y-point were implicitly assumed.
  Therefore, the fact that the only stable configuration which was
  found in these simulations was the configuration with $R_Y=\RLC$
  does not exclude possibility that other stable configurations with
  different $\Omega_F$ and $R_Y$ exist.},
results in a set of meta-stable magnetospheric configurations, in the
sense that such configurations represent local minima of the total
energy -- any small deviations of the current density and/or the size
of the corotating zone from that in a given ``minimal'' configuration
would result in a larger value of the total energy.

Changes in the current density distribution require variations of the
potential drop in the polar cap.  These variations should be of the
order of several percent of the potential drop across the polar cap
in order to produce noticeable changes in $\Omega_F$
\citep{Timokhin2007:MNRAS2}.  In young pulsars, where a very small
fraction of the vacuum potential drop would be enough to short out the
accelerating electric field, these variations should be small.  With
aging of a pulsar, the freedom in the current density distribution will
increase, thus making existence of multiple metastable configurations
of the magnetosphere possible.  In the frame of the proposed model
this might be the reason why mode changes and nulling are seen in
older pulsars.

The pulsar magnetosphere represents a case of a highly non-linear
system with a complicated means of adjustment between its
subcomponents.  For example, the polar cap cascade zone should adjust
the current density flowing through it to that required by the
magnetosphere.  It is not clear now how this adjustment works.  The
characteristic timescale of the cascade zone is of the order of
microseconds, while the magnetospheric characteristic timescale is of
the order of the pulsar rotation period.  Pulsar magnetosphere being
highly non-linear system with very different characteristic time
scales might exhibit rich dynamical behavior.  In particular, some of
these meta-stable configurations may be a sort of strange attractors,
where a system spends substantial time and then suddenly changes to a
different state.  The time intervals between these changes may be very
large, much larger than any characteristic time scale of the system
and transitions can happen quasi-periodically.  This might account for
quasi-periodicity in changes between ON and OFF states in PSR
B1931+24.

Rotating radio transients (RRATs) -- recently discovered transient
radio sources -- are thought to be rotating neutron stars too
\citep{McLaughlin2006}.  The discovery of pulsar PSR J0941-39 which
switches between a RRAT-like isolated burst-emitting mode and a mode
with emission typical for a nulling pulsar might imply that RRATs are
an extreme case of nulling pulsars \citep{Burke-Spolaor2010}.  In the
frame of the proposed model it means that RRATs are nulling pulsars
which spend most of their time in a state(s) where our line of sight
does not cross the beam or no radioemission is generated.  The
existence of such pulsars seems to be a natural consequence of our
model.

As a direct hint that the pulsar magnetosphere can indeed evolve on
timescales much larger than the rotational period, we can consider the
evolution of subpulse drift rate in PSR B0943+10
\citep{RankinSuleymanova2006}.  In this mode-changing pulsar the
subpulse drift rate evolves after the offset of the ``B'' mode with a
characteristic timescale of some $\sim4000$ rotational periods.  This
time is larger that any characteristic timescale the force-free
magnetosphere can have.  It might be that in this case we directly
observe manifestations of some non-linear process(es) involving
adjustment between the magnetosphere, the polar cap cascades and/or
the current sheet.  If the pulsar magnetosphere is indeed a highly
non-linear system, then the existence of different metastable states
seems to be a quite natural assumption.

\section*{Acknowledgments}

I wish to thank Jonathan Arons and Joeri van Leeuwen for fruitful
discussions.  I am grateful to the referee Joanna Ranking for very
helpful comments.  This work was supported by NSF grant AST-0507813,
NASA grants NNG06GJI08G, NNX09AU05G and DOE grant DE-FC02-06ER41453.

\bibliographystyle{mn2e} 

\bibliography{/home/atim/ARTICLES/Bibliographies/pulsars/pulsars_theory,/home/atim/ARTICLES/Bibliographies/pulsars/pulsars_obs,/home/atim/ARTICLES/Bibliographies/magnetars/magnetars,/home/atim/ARTICLES/Bibliographies/NumericalMethods/numerical_methods}

\begin{thebibliography}{}

\bibitem[\protect\citeauthoryear{{Abdo} \& et al.}{{Abdo} \&
  et~al.}{2010}]{FERMI_PSR_Catalog2010}
{Abdo} A.~A.,  et al. 2010, \apjs, 187, 460

\bibitem[\protect\citeauthoryear{{Arons}}{{Arons}}{1983}]{Arons1983a}
  {Arons} J., 1983, in Burns M.~L., Harding A.~K., Ramaty R., eds, AIP
  Conf. Ser., Vol.~101, Positron-Electron Pairs in Astrophysics.
  Am. Inst. Phys., New York, p. 163

\bibitem[\protect\citeauthoryear{{Biggs}}{{Biggs}}{1992}]{Biggs1992}
{Biggs} J.~D.,  1992, \apj, 394, 574

\bibitem[\protect\citeauthoryear{{Blandford} \& {Znajek}}{{Blandford} \&
  {Znajek}}{1977}]{Blandford/Znajek77}
{Blandford} R.~D.,  {Znajek} R.~L.,  1977, \mnras, 179, 433

\bibitem[\protect\citeauthoryear{{Bucciantini}, {Thompson}, {Arons}, {Quataert}
  \& {Del Zanna}}{{Bucciantini} et~al.}{2006}]{Bucciantini06}
{Bucciantini} N.,  {Thompson} T.~A.,  {Arons} J.,  {Quataert} E.,    {Del
  Zanna} L.,  2006, \mnras, 368, 1717

\bibitem[\protect\citeauthoryear{{Burke-Spolaor} \& {Bailes}}{{Burke-Spolaor}
  \& {Bailes}}{2010}]{Burke-Spolaor2010}
{Burke-Spolaor} S.,  {Bailes} M.,  2010, \mnras, 402, 855

\bibitem[\protect\citeauthoryear{{Contopoulos}}{{Contopoulos}}{2005}]{Contopou%
los05}
{Contopoulos} I.,  2005, \aap, 442, 579

\bibitem[\protect\citeauthoryear{{Contopoulos}, {Kazanas} \&
  {Fendt}}{{Contopoulos} et~al.}{1999}]{CKF}
{Contopoulos} I.,  {Kazanas} D.,    {Fendt} C.,  1999, \apj, 511, 351

\bibitem[\protect\citeauthoryear{{Contopoulos} \& {Spitkovsky}}{{Contopoulos}
  \& {Spitkovsky}}{2006}]{Contopoulos2006}
{Contopoulos} I.,  {Spitkovsky} A.,  2006, \apj, 643, 1139

\bibitem[\protect\citeauthoryear{{Goldreich} \& {Julian}}{{Goldreich} \&
  {Julian}}{1969}]{GJ}
{Goldreich} P.,  {Julian} W.~H.,  1969, \apj, 157, 869

\bibitem[\protect\citeauthoryear{Gruzinov}{Gruzinov}{2005}]{Gruzinov:PSR}
Gruzinov A.,  2005, Phys.Rev.Lett., 94, 021101

\bibitem[\protect\citeauthoryear{{Kalapotharakos} \&
  {Contopoulos}}{{Kalapotharakos} \& {Contopoulos}}{2009}]{Kalapotharakos2009}
{Kalapotharakos} C.,  {Contopoulos} I.,  2009, \aap, 496, 495

\bibitem[\protect\citeauthoryear{{Kramer}, {Lyne}, {O'Brien}, {Jordan} \&
  {Lorimer}}{{Kramer} et~al.}{2006}]{Kramer2006}
{Kramer} M.,  {Lyne} A.~G.,  {O'Brien} J.~T.,  {Jordan} C.~A.,    {Lorimer}
  D.~R.,  2006, Science, 312, 549

\bibitem[\protect\citeauthoryear{{Lyne}}{{Lyne}}{2009}]{Lyne2009}
  {Lyne} A.~G., 2009, in Becker W., ed., Astrophysics and Space
  Science Library, Vol.~357, Neutron Stars and Pulsars. Springer
  Berlin p.~67

\bibitem[\protect\citeauthoryear{{McLaughlin}, {Lyne}, {Lorimer},
    {Kramer}, {Faulkner}, {Manchester}, {Cordes}, {Camilo},
    {Possenti}, {Stairs}, {Hobbs}, {D'Amico}, {Burgay} \&
    {O'Brien}}{{McLaughlin} et~al.}{2006}]{McLaughlin2006}
  {McLaughlin} M.~A. et~al., 2006, \nat, 439, 817

\bibitem[\protect\citeauthoryear{{Michel}}{{Michel}}{1973}]{Michel73}
{Michel} F.~C.,  1973, \apjl, 180, L133

\bibitem[\protect\citeauthoryear{{Rankin} \& {Suleymanova}}{{Rankin} \&
  {Suleymanova}}{2006}]{RankinSuleymanova2006}
{Rankin} J.~M.,  {Suleymanova} S.~A.,  2006, \aap, 453, 679

\bibitem[\protect\citeauthoryear{{Spitkovsky}}{{Spitkovsky}}{2006}]{Spitkovsky%
:incl:06}
{Spitkovsky} A.,  2006, \apjl, 648, L51

\bibitem[\protect\citeauthoryear{{Timokhin}}{{Timokhin}}{2006}]{Timokhin2006:M%
NRAS1}
{Timokhin} A.~N.,  2006, \mnras, 368, 1055

\bibitem[\protect\citeauthoryear{{Timokhin}}{{Timokhin}}{2007a}]{Timokhin2007:%
MNRAS2}
{Timokhin} A.~N.,  2007a, \mnras, 379, 605

\bibitem[\protect\citeauthoryear{{Timokhin}}{{Timokhin}}{2007b}]{timokhin::PSR%
EQ_london/2006}
{Timokhin} A.~N.,  2007b, \apss, 308, 575

\bibitem[\protect\citeauthoryear{{Wang}, {Manchester} \& {Johnston}}{{Wang}
  et~al.}{2007}]{Wang2007}
{Wang} N.,  {Manchester} R.~N.,    {Johnston} S.,  2007, \mnras, 377, 1383

\end{thebibliography}

\label{lastpage}

\end{document}